*Strain coupling of a mechanical resonator to a single quantum emitter in diamond*

Kenneth W. Lee[1, †], Donghun Lee[1,2, †], Preeti Ovartchaiyapong[1], Joaquin Minguzzi[3], Jero R. Maze[3], and Ania C. Bleszynski Jayich[1,*]

**Abstract**

The recent maturation of hybrid quantum devices has led to significant enhancements in the functionality of a wide variety of quantum systems. In particular, harnessing mechanical resonators for manipulation and control has expanded the use of two-level systems in quantum information science and quantum sensing. In this letter, we report on a monolithic hybrid quantum device in which strain fields associated with resonant vibrations of a diamond cantilever dynamically control the optical transitions of a single nitrogen-vacancy (NV) defect center in diamond. We quantitatively characterize the strain coupling to the orbital states of the NV center, and with mechanical driving, we observe NV-strain couplings exceeding 10 GHz. Furthermore, we use this strain-mediated coupling to match the frequency and polarization dependence of the zero-phonon lines of two spatially separated and initially distinguishable NV centers. The experiments demonstrated here mark an important step toward engineering a quantum device capable of realizing and probing the dynamics of non-classical states of mechanical resonators, spin-systems, and photons.

[1] Department of Physics, University of California Santa Barbara, Santa Barbara, CA 93106

[2] Department of Physics, Korea University, Seoul, South Korea

[3] Institute of Physics, Pontificia Universidad Católica de Chile, Santiago 7820436, Chile

[†] These authors contributed equally to this work

[*] e-mail: ania@physics.ucsb.edu

# I. INTRODUCTION

Hybrid quantum devices involving mechanical systems are attractive candidates for future quantum technologies. Mechanical systems are sensitive to a diverse array of external forces, and with recent advancements made in nanofabrication techniques, these systems can be precisely engineered with high quality factors. Several experiments have exploited this sensitivity to engineer magnetic, electric, and strain couplings between mechanical resonators and two-level systems in atoms, quantum dots, superconducting Josephson junctions and defects in semiconductors [1-7]. These experiments have demonstrated coherent mechanical control of two-level systems and in some cases, coupling of the two-level system to the quantum motion of the resonator [1]. Ultimately, these devices strive to address several outstanding challenges in the development of quantum technologies, including the realization of long-range interactions between a wide variety of quantum systems, and the preparation of non-classical states of a macroscopic mechanical object.

A hybrid spin-mechanical device where the coupling is mediated by crystal strain offers several promising applications in the fields of quantum information science and sensing. For example, crystal strain associated with the spatially extended phonon modes of mechanical resonators can be used as a quantum data bus to transmit quantum information between disparate qubits [8,9] , prepare spins in entangled or squeezed states [10], or provide an interface between photons of disparate energy scales [11,12]. This interaction can similarly be used to cool mechanical resonators to their quantum ground state [13,14], enabling fundamental studies of quantum mechanics in macroscopic objects [1,15,16].

From a practical standpoint, using crystal strain to interface a mechanical resonator with a two-level system has many advantages. Strain directly couples the two systems without the need for external components or functionalization of the resonator, and therefore allows for a monolithic architecture. Importantly, the coupling does not generate noisy stray fields, which may degrade the coherence of the system. Furthermore, a monolithic architecture eliminates drifts in the coupling strength and allows for more facile device fabrication, which is crucial when scaling the device to large numbers of quantum systems.

A leading candidate for a strain-coupled hybrid device consists of nitrogen-vacancy (NV) defect center spins and diamond mechanical resonators [17-20]. The electron spin of the NV center has been shown to exhibit long coherence times near one second at 77 K [21] and single-crystal diamond mechanical resonators have been fabricated with high quality factors in excess of one million [22]. To date, experiments coupling mechanical oscillators to the NV spin have realized mechanically driven spin dynamics, enabling coherent spin manipulation in the strong coupling regime [23], nanoscale strain sensing [17], and continuous dynamical decoupling from magnetic field noise [23,24]. However, proposed hybrid quantum devices involving the NV center, photons, and phonons will also require a mechanical coupling to the orbital states of the NV center [25]. The orbital degree of freedom is inherently more sensitive to crystal strain than the spin degree of freedom, and thus provides higher single-phonon couplings. In addition, coupling to the orbital states allows for direct tuning of the optical transitions of the NV center, which is essential for photonic applications that require generation of indistinguishable photons, such as entanglement of distant spins [26]. Additionally, strain-orbit coupling enables several applications that are inaccessible with spin-strain coupling, such as phonon cooling of a mechanical resonator [14] and phonon routing [25]. Very recently, coherent control of the NV orbital states was demonstrated using a combination of optical fields and strain fields associated with propagating surface acoustic waves generated by an interdigitated transducer [27].

In this letter, we realize a dynamic, strain-mediated coupling between the orbital states of a single NV center in diamond and the resonant mechanical motion of a diamond cantilever. The time-varying strain field generated by the cantilever displacement modulates both the energy and polarization dependence of the optical transitions of the NV center, which we characterize using resonant excitation spectroscopy (RES). With mechanical driving, we observe large (10 GHz) NV strain-orbit coupling and precisely characterize the strain-orbit coupling constants. Notably, we observe strain-mediated single phonon coupling strengths of a few kHz, which are three to five orders of magnitude higher than those previously demonstrated with the NV spin [17-20]. Finally, we use this coupling to match the energy and polarization dependence of the zero-phonon lines of two spatially separated NV centers.

## II.     EXPRIMENTAL SETUP AND THEORETICAL MODEL

Our experiments are performed on single, negatively-charged NV centers embedded in a single-crystal diamond cantilever as depicted in Fig. 1a. In Fig. 1c, we show a simplified energy level diagram of the NV center that depicts the relevant energy levels investigated here. An orbital-singlet/spin-triplet ground state, $^3A_2$, is connected to an orbital-doublet/spin-triplet excited state, $^3E$, by a zero-phonon dipole transition at 637 nm. Here, we focus on electronic states with spin projection $m_s = 0$, consisting of two excited state levels ($|E_x\rangle, |E_y\rangle$) and one ground state level ($|A\rangle$) (although $|E_x\rangle$ and $|E_y\rangle$ are slightly mixed via spin-spin interaction with $m_s = \pm 1$ projections under perfect $C_{3v}$ symmetry, we neglect this effect and approximate them as the $m_s = 0$ spin projections of the excited state). Experiments are carried out at cryogenic temperatures (~7 K) where phonon broadening of the optical transitions is negligible and the excited state fine structure is accessible [28]. We use a home-built confocal microscope to optically address single NV centers in diamond cantilevers and to interferometrically monitor the cantilevers' motion. In our experiments, we off-resonantly excite the NV center with 532 nm laser light to identify NV centers in the cantilever (Fig. 1b), initialize them into the $m_s = 0$ spin sublevel of the ground state, and to stabilize the NV$^-$ charge state. To perform RES of the $A \rightarrow E_x$ and $A \rightarrow E_y$ transitions, we use 637 nm linearly polarized light from a tunable diode laser and monitor the NV photoluminescence (PL). A piezoelectric transducer located below the sample holder mechanically actuates the cantilever, and a 450 nm laser is used for interferometric detection of the cantilever's motion.

Coupling of the cantilever motion to the NV orbital states is provided by the sensitivity of the NV center to crystal strain. Strain deforms the molecular orbitals of the NV center, and its effect on the orbital levels is determined by the NV's $C_{3v}$ symmetry [29]. In the $\{|A\rangle, |E_x\rangle, |E_y\rangle\}$ basis, the strain-coupled NV-resonator Hamiltonian ($h=1$) takes the form

$$H_{NV} = f_{ZPL}\left(|E_x\rangle\langle E_x| + |E_y\rangle\langle E_y|\right) + \frac{\omega_c}{2\pi}a^\dagger a$$
$$+ \left[g_{A_1}\left(|E_x\rangle\langle E_x| + |E_y\rangle\langle E_y|\right) + g_{E_1}\left(|E_x\rangle\langle E_x| - |E_y\rangle\langle E_y|\right) + g_{E_2}\left(|E_x\rangle\langle E_y| + |E_x\rangle\langle E_y|\right)\right]\left(a+a^\dagger\right)$$

(1)

where $f_{ZPL}$ is the natural zero-phonon line frequency, $\frac{\omega_c}{2\pi}$ is the cantilever frequency, $g_\Gamma$ is the single phonon coupling strength for phonons with symmetry $\Gamma$, $a^\dagger(a)$ are the creation (annihilation) operators for the phonon mode of the cantilever, and we have defined $|A\rangle$ to be at zero energy. In a perfect crystal lattice with no applied external fields, the NV has perfect $C_{3v}$ symmetry and the $|E_x\rangle$ and $|E_y\rangle$ orbital levels are degenerate in energy. Strain of $A_1$ symmetry corresponds to dilations and contractions of the NV molecular structure that preserve the $C_{3v}$ symmetry. An $A_1$-symmetric strain preserves the $E_x$ and $E_y$ degeneracy and uniformly shifts the two states in energy with respect to $|A\rangle$. Conversely, strains of $E_1$ and $E_2$ symmetry break all symmetries of the NV center, and therefore split and mix the $E_x$ and $E_y$ states respectively (Fig 1c). From the NV-resonator Hamiltonian, the $E_x$ and $E_y$ transition frequencies, $f_+$ and $f_-$, can be expressed in terms of the normalized cantilever displacement, $x = x_c/x_0$,

$$f_\pm(x) = f_{ZPL} + \delta f_{A_1} + g_{A_1} x \pm \sqrt{\left(g_{E_1} x + \delta f_{E_1}\right)^2 + \left(g_{E_2} x + \delta f_{E_2}\right)^2} \qquad (2)$$

where $x_0$ is the amplitude of zero-point motion, $x_c$ is the peak displacement of the cantilever, and $\delta f_\Gamma$ corresponds to the frequency shift due to local, intrinsic strain of symmetry $\Gamma$.

### III. EXPERIMENTAL RESULTS AND DISCUSSION

#### A. DYNAMIC STRAIN-ORBIT COUPLING

We demonstrate strain-mediated NV-phonon coupling by mechanically driving the fundamental flexural mode of a cantilever with a resonance frequency $\omega_c/2\pi = 870$ kHz and performing RES on a single embedded NV (Fig. 2). In the absence of mechanical driving ( we observe two Lorentzian peaks in the RES spectrum (gray) corresponding to the $E_x$ and $E_y$ transitions. Under a resonant mechanical drive the peaks are modulated (green) by the AC strain field produced by the cantilever's flexural motion, which is given by $x(t) = x_c/x_0 \cos(\omega_c t)$. The broadened, double-peaked lineshape results from averaging over the cantilever motion and can be modeled by a Lorentzian with an oscillating center frequency as defined in equation (2).

In Fig. 2b, we show the dynamic modulation of the $E_x$ and $E_y$ transitions as the drive frequency is swept through the cantilever resonance. The resonant behavior of the mechanical mode is clearly imprinted on the spectral response of the NV center. Far detuned from the mechanical resonance, the transitions are unaffected and maintain their Lorentzian lineshape. Approaching resonance, the lineshapes broaden due to the increasing deflection of the cantilever. The spectral response of the NV is in good agreement with simulations shown in Fig. 2c, which take into account the intrinsic strain environment and the mechanical quality factor of 20,000 (see Appendix D). The asymmetric frequency modulation of the optical transitions, most clearly seen at zero mechanical detuning, arises from intrinsic crystal strain of $E_1$ and $E_2$ symmetry, which we discuss in greater detail later. In addition, the intrinsic $E_1$- and $E_2$- symmetric strains are responsible for the splitting between the $|E_x\rangle$ and $|E_y\rangle$ when the cantilever is at rest,

$$\Delta f_0 = 2\sqrt{\left(\delta f_{E_1}\right)^2 + \left(\delta f_{E_2}\right)^2}.$$

Next, we probe the strength of the strain-mediated NV-phonon coupling as a function of cantilever deflection for a resonant mechanical drive (Fig. 2d). As the amplitude of beam deflection increases, the strain-modulation of the optical transitions increases approximately linearly, as predicted by equation (2). For the NV measured in Fig. 2, we observe strain-mediated NV-phonon couplings exceeding 10 GHz for cantilever displacements of only a few nanometers. Our measurement is in good agreement with simulations (Fig. 2e) and demonstrates the large coupling of the orbital states to the mechanical motion of the cantilever. We note that the observed coupling strengths are limited by the scanning range of the resonant excitation laser and not by the device itself.

## B. MEASUREMENT OF THE STRAIN-ORBIT COUPLING CONSTANTS

To accurately determine the single-phonon coupling parameters, we must isolate the effects of strain for each symmetry, which cannot be accomplished with the continuous wave (CW) spectroscopy technique employed above. The CW spectral response of the NV is a time-averaged response that samples all points of the cantilever's motion, making it difficult to distinguish between common mode shifts and splittings of the $|E_x\rangle$ and $|E_y\rangle$ states. However, by performing stroboscopic RES synchronized to the cantilever's motion, we can deconvolve the orbital dynamics from the mechanical motion [30]. Specifically, we measure the NV fluorescence in 60 ns windows that are much shorter than the cantilever's 1.15 μs oscillation period, thus enabling RES at a well-defined cantilever position. In Fig. 3a, we demonstrate stroboscopic RES when the cantilever is maximally deflected upward (blue) and maximally deflected downward (orange) with an amplitude of 24 nm. Importantly, we can distinguish the effects of uniform shifts of the transitions due to $A_1$ strain from splittings due to $E$ strain, and with the well-known strain profile of a cantilever, we may extract the orbital strain coupling constants, $\lambda_{A_1}$, $\lambda_{A_1'}$, $\lambda_E$, and $\lambda_{E'}$ [16,17]. The single-phonon coupling parameters can be written explicitly in terms of the orbital-strain coupling constants and the strain induced by the zero-point motion of the cantilever, $\epsilon$:

$$g_{A_1} = \lambda_{A_1}\epsilon_{zz} + \lambda_{A_1'}(\epsilon_{xx} + \epsilon_{yy}), \qquad g_{E_1} = \lambda_E(\epsilon_{yy} - \epsilon_{xx}) + \lambda_{E'}(\epsilon_{xz} + \epsilon_{zx}), \quad \text{and}$$

$g_{E_2} = \lambda_E(\epsilon_{xy} + \epsilon_{yx}) + \lambda_{E'}(\epsilon_{yz} + \epsilon_{zy})$ (see Appendix B).

In Fig. 3b, we study the $A_1$-symmetric common mode shifts of the $|E_x\rangle$ and $|E_y\rangle$ levels for 12 different NV centers as a function of the mechanically induced strain along the cantilever axis, which lies along the [110] crystal direction. We observe two distinct spectral responses, a result of probing NVs of different crystal orientations. The NVs fall into two groups: group A with NVs oriented $[\bar{1}\bar{1}\bar{1}]$ and $[11\bar{1}]$ and group B with NVs oriented $[\bar{1}11]$ and $[1\bar{1}1]$. Within each group, NVs experience the same type of strain induced by the cantilever motion (see Appendix H). Fitting this data, we find $\lambda_{A_1} = -1.95 \pm 0.29 \, \text{PHz}$ and $\lambda_{A_1'} = 2.16 \pm 0.32 \, \text{PHz}$. To extract the two remaining coupling constants, we analyze the total frequency shifts of the $E_x$ and $E_y$ transitions for NVs in groups A (Fig. 3c) and B (Fig. 3d). To accurately fit the data, we must take into account the intrinsic $E$-symmetric strain environment of each NV center that generates $\delta f_{E_1}$ and $\delta f_{E_2}$. $E$-symmetric strain rotates the orthogonal transition dipole moments for the $E_x$ and $E_y$ transitions, and thus modifies the polarization dependence of the transitions. The rotation angle, $\theta$, is defined by $\tan(2\theta) = \left(\delta f_{E_2} + g_{E_2} x\right) / \left(\delta f_{E_1} + g_{E_1} x\right)$, and is defined relative to the NV $x$-axis (see Appendix I). Measuring both $\Delta f_0$ and $\theta$ allows for complete characterization of the intrinsic strain environment of an NV center. From the fits of the transition frequencies in Fig. 3c-d, we obtain $\lambda_E = -0.85 \pm 0.13 \, \text{PHz}$ and $\lambda_{E'} = 0.02 \pm 0.01 \, \text{PHz}$. From the 12 NVs measured in this paper, we observe maximal single phonon couplings of $g_{A_1} \sim 1 \, \text{kHz}$ and $g_{E_1} \sim 3 \, \text{kHz}$, and note that due to the crystallographic orientation of the cantilever, $g_{E_2} = 0$.

## C. STROBOSCOPIC FREQUENCY TUNING OF A QUANTUM EMITTER

A fundamental challenge for engineering photon-mediated interactions between solid-state quantum emitters is the generation of indistinguishable photons [26,31]. Solid-state quantum emitters are hosted in a crystal lattice and experience locally varying electrostatic and strain environments, resulting in inhomogeneous optical emission properties. Previous experiments have addressed this obstacle using multi-axis DC electric fields applied by electrodes near or on the diamond sample [32,33]. However, electric field tuning can exacerbate spectral diffusion due to disturbances of the local electrostatic environment and electric field noise from the electrodes, degrading photon indistinguishability. Here, we take a promising alternative approach that uses our hybrid strain-mediated coupling to control the frequency of the $E_x$ and $E_y$ transitions. This control will allow for dynamic matching of the optical transition frequencies of two spatially separated NV centers. For weakly or non-piezoelectric host materials such as diamond, strain-control should introduce minimal spectral diffusion to embedded quantum emitters.

Figs. 4a and c show RES measurements of two NVs: one located in the cantilever (NV I) and one located in the bulk of our sample (NV II). In the absence of mechanical driving, their $E_x$ and $E_y$ transitions differ by several GHz. By resonantly driving the cantilever, we can tune the $E_x$ and $E_y$ transitions of NV I into resonance with NV II. Fig. 4b shows stroboscopic RES measurements of NV I for various cantilever deflection amplitudes. At amplitudes of 43 nm and 30 nm respectively, the $E_x$ and $E_y$ transition frequencies of NV I match those of NV II. Importantly, this technique introduces very minimal spectral broadening, which can be attributed to the finite duration of the detection window. In the future, this technique could be extended to a large array of cantilevers, each with a different resonant frequency, enabling on-chip, multiplexed tuning of many NV centers for multipartite entanglement. In this architecture, a high density of qubits can be addressed without crosstalk between photonic channels using nanoscale cantilevers without the need for individual electrical leads for each NV. Excitation of multiple resonators is straightforward and can be carried out by driving a single piezoelectric transducer at multiple frequencies. This AC strain tuning modality is also more tractable than protocols involving static strain, which may suffer from an unknown strain profile and may require significantly more fabrication, alignment, and stabilization steps to individually address each structure. In addition, strain tuning could be used for dynamic feedback control of the transitions to reduce spectral diffusion caused by local charge fluctuations [33].

### D. STROBOSCOPIC POLARIZATION TUNING OF A QUANTUM EMITTER

The polarization dependence of the optical transitions of spatially separated NV centers can also be matched using our strain-mediated control scheme. We characterize the polarization dependence by performing RES for varying linear laser polarization angles (Fig. 5a) and measuring the corresponding PL amplitudes of the transition peaks. In Figs. 5b-e, we plot the polarization dependence of the $E_x$ and $E_y$ transitions for 4 different NV centers, where NVs A, B, and C are located in the sample bulk, and NV D is located inside the cantilever. Due to their inhomogeneous environments, they all exhibit different polarization dependences, with $\theta = (16.4°, -134.1°, -116.4°, 33.9°)$ for NVs A, B, C and D, respectively. In Figs. 5f-h, we match the polarization dependence of the $E_x$ and $E_y$ transitions of NV D to those of NV A, B and C by driving the cantilever at tip amplitudes of 9.4 nm, 3.7 nm, and 12.4 nm respectively, and measuring when the cantilever is maximally deflected downward. The dashed curves in Figs. 5f-h correspond to the expected polarization dependence, taking into account the cantilever amplitude and distortions in the shape of the dipole excitation patterns caused by non-zero ellipticity in the laser beam and saturation of the optical transitions. The excellent agreement between measured and expected polarization dependences highlights the deterministic nature of our strain-control.

*In-situ* polarization tuning is an important and practical function for future hybrid devices that incorporate quantum emitters. Typically, the polarization state of spontaneously emitted photons is manipulated after the collection optics with a combination of polarization control elements, such as waveplates and fiber paddles. For fully chip-integrated systems where many quantum emitters need to be individually addressed, it can be difficult to utilize these external components for selective control of a photonic channel. *In-situ* control allows for selective control while minimizing cross-talk between photonic channels. In addition to strain, *in-situ* control can be accomplished with electric fields [34] or optical microcavities [35]. However, the fabrication and alignment required for individual addressing with electric fields and cavities may present a significant challenge as the number of emitters increases.

In our current resonator geometry we do not have independent control of $E_1$ and $E_2$ strain, and are thus unable to arbitrarily tune the frequency and polarization dependence of an optical transition simultaneously. Universal control could be enabled by simultaneous excitation of two mechanical modes that generate different strain profiles. Interestingly, this approach could offer novel opportunities to study multimode optomechanics by utilizing the inherent nonlinearity of a two level system [36]. Alternatively, generation of indistinguishable photons can still be accomplished by a combination of frequency matching and polarization filtering [37,38].

## IV. CONCLUSION

Future quantum-enabled applications such as long-range interactions between distant qubits or phonon routing will require the single phonon coupling strength to be larger than the intrinsic dissipation in the hybrid system [8,10,25]. This high cooperativity regime is defined by $\eta = g^2/\Gamma_2\gamma_{th} > 1$, where $\eta$ is the cooperativity, $\Gamma_2$ is the dephasing rate of the optical transition and $\gamma_{th}$ is the thermalization rate of the resonator with the environment. By scaling our device down to the nanoscale and improving optical linewidths, it should be possible to enter the high cooperativity regime. For instance, the fundamental flexural mode of a doubly-clamped diamond beam of dimensions $2\,\mu m \times 100\,nm \times 50\,nm$ with $\omega_c/2\pi = 230\,\text{MHz}$ and $Q = 10^5$, coupled to a near surface NV center with $\Gamma_2 = 100\,\text{MHz}$, would have a coupling $g = 21.5\,\text{MHz}$ and a cooperativity $\eta \sim 5$ at $T = 4\,K$, residing deep within the high cooperativity regime. Moreover, this device could be used for single-defect cooling of the fundamental mode of the nanobeam to its quantum ground state [13,14] with phonon occupation number $\bar{n} < 1$, allowing for preparation and observation of quantum states of the mechanical oscillator (see Appendix K).

In summary, we have presented a hybrid quantum device in which the orbital states of a single solid-state quantum emitter are coupled via strain to a single vibrational mode of a high quality factor mechanical resonator. We used our device to carefully control the frequency and polarization dependence of the optical transitions of single NV centers. Several additional and diverse applications could be enabled by our device. For instance, our hybrid coupling mechanism could be used to generate large, broadly-tunable light phase shifts, enabling novel applications in all-optical signal processing [39]. Moreover, by combining our strain-mediated NV-phonon coupling with electron spin resonance at the cryogenic temperatures employed here, it should be possible to significantly enhance the optically detected magnetic resonance signal through phonon assisted optical pumping [40]. Additionally, our dynamic strain-tuning experiments may be extended for manipulation and control of the NV hyperfine transitions [41], which may facilitate the use of nuclear spins as quantum memories [42]. Furthermore, our device architecture and its applications are compatible with other solid-state emitters, such as the silicon-vacancy center in diamond or defects in silicon carbide [43]. Most importantly, we have demonstrated for the first time strain-coupling of a mechanical oscillator to the orbital states of the NV center. Scaling down our device dimensions and reducing optical linewidths should allow for operation in the resolved sideband regime, enabling phonon cooling and lasing [14], spin-dependent force quantum gates [10], and phononic routing applications [9,25]. The single phonon couplings shown here are the largest ever demonstrated for an NV-based system, and constitute a major step toward quantum applications enabled by a hybrid quantum device based on spins, phonons and photons.

## ACKNOWLEDGEMENTS

We thank Matt Pelliccione and Alec Jenkins for experimental assistance. This work is supported by the Air Force Office of Scientific Research Quantum Memories MURI program, an NSF-CAREER award (DMR-1352660), a Fondecyt-Conicyt grant (No. 1141185), and an AFOSR grant (FA9550-15-1-0113).

# APPENDIX A: EXPERIMENTAL SETUP AND SAMPLE PREPARATION

The experimental setup consists of a homebuilt cryogenic confocal microscope as shown in Fig. 6a. Experiments are carried out at 6.7 K in a closed-cycle cryostat (Montana Instruments). A continuous wave diode laser (Laser Quantum) at 532 nm is used to optically pump the NV center into the $m_s = 0$ spin level of the ground state, and to initialize the negative charge state of the NV. A tunable diode laser near 637 nm (New Focus Velocity) is used for resonant excitation of the NV zero-phonon line, and is gated with an acousto-optic modulator (Intraaction). A combination of linear polarizers and a liquid-crystal retarder is used to control the linear polarization of the resonant excitation light. The relative polarization of the laser light with respect to the diamond crystallographic orientations is monitored with a polarimeter (Thorlabs). Photons emitted by the NV are collected into a single-mode fiber and subsequently detected by an avalanche photodiode (APD). A continuous wave diode laser at 450 nm (Thorlabs) is used for interferometric detection of the mechanical motion of the cantilever. The mechanical motion of the cantilever phase modulates the reflected light off the cantilever surface and interferes with light that passes through the cantilever and reflects off the silicon substrate 1 μm below the cantilever. The 450 nm light is filtered out of the collection path using a longpass dichroic filter (Semrock). A piezoelectric transducer with a capacitance of 17 nF is mounted just below the sample for mechanical driving of the diamond cantilevers.

The cantilever used in this experiment was fabricated using the technique described in ref. [17]. NV centers were formed via nitrogen implantation, with a dosage of $3 \times 10^9$ ion/cm² at an energy of 40 keV and a 0° tilt, yielding an expected depth of 51.5 nm (calculated by Stopping and Range of Ions in Matter (SRIM)). The sample was annealed under high vacuum ($10^{-6}$ Torr) at 800 °C for 3 hours.

# APPENDIX B: THE STRAIN INTERACTION AND SINGLE-PHONON COUPLING PARAMETERS

The internal degrees of freedom of the NV center are sensitive to crystal strain through the ion-electron Coulomb interaction. The strain-induced displacement of ions in the diamond lattice alters the Coulomb potential felt by the electrons associated with the NV center. This directly affects the orbital degree of freedom of the NV center. We note that spin-orbit coupling allows for strain to couple to the spin degree of freedom as a higher order perturbation.

To analyze the effect of crystal strain on the NV, let us consider a generic strain experienced by the NV center as defined in its own coordinate basis, which is characterized by the strain tensor $\epsilon$:

$$\epsilon = \begin{bmatrix} \epsilon_{xx} & \epsilon_{xy} & \epsilon_{xz} \\ \epsilon_{yx} & \epsilon_{yy} & \epsilon_{yz} \\ \epsilon_{zx} & \epsilon_{zy} & \epsilon_{zz} \end{bmatrix} \qquad (3)$$

The NV's coordinate system is shown in Fig. 7. The $z$-axis, or NV axis, lies along the nitrogen bond, and can point in 4 crystallographic directions: $[\bar{1}\bar{1}\bar{1}]$, $[11\bar{1}]$, $[\bar{1}11]$ or $[1\bar{1}1]$. The NV $x$-axis lies along the projection of a carbon bond in the plane perpendicular to the NV axis. The strain Hamiltonian can be written in terms of individual components of the strain tensor in the following way [29]

$$H_{strain} = \sum_{i,j} \Lambda_{ij} \epsilon_{ij} \quad (4)$$

where $i,j$ refer to Cartesian indices of the NV coordinate system and $\Lambda_{ij}$ are orbital operators. The NV center has $C_{3v}$ symmetry, and hence it is useful to project the strain Hamiltonian onto the irreducible representations of the $C_{3v}$ group [29].

$$H_{strain} = \sum_{\Gamma} \Lambda_{\Gamma} \epsilon_{\Gamma} \quad (5)$$

In the above Hamiltonian we take advantage of the fact that since strain is a symmetric second-rank tensor, its components transform as the quadratic basis functions of the $C_{3v}$ group. Therefore, $H_{strain}$ will only contain components that transform as $A_1$ and $E$. Hence, $\Gamma$ indexes the $A_1$ and $E$ irreducible representations of the $C_{3v}$ group. For instance, $\epsilon_{zz}$ transforms like $z^2$, and hence it will transform as $A_1$. Strains that transform as $A_1$ will preserve the $C_3$ rotational symmetry and $\sigma_v$ reflection symmetry of the NV center; whereas, strains that transform as $E$ will break both the rotation and reflection symmetries.

The NV center consists of an orbital-singlet/spin-triplet ground state, $^3A_2$, that is connected to an orbital-doublet/spin-triplet excited state $^3E$ by an optical zero-phonon line. In this paper, we focus on the $m_s = 0$ spin levels of the $^3A_2$ and $^3E$ manifolds. We label the single $m_s = 0$ level of the ground state as $|A\rangle$ and the two excited state levels as $|E_x\rangle$ and $|E_y\rangle$.

In the $\{|A\rangle, |E_x\rangle, |E_y\rangle\}$ basis where $|A\rangle$ defines zero energy, the NV Hamiltonian including strain can be written as

$$H_{strain} = \left[ f_{ZPL} + \lambda_{A_1} \epsilon_{zz} + \lambda_{A_1'} \left( \epsilon_{xx} + \epsilon_{yy} \right) \right] \left[ |E_x\rangle\langle E_x| + |E_y\rangle\langle E_y| \right] +$$

$$+ \left[ \lambda_E \left( \epsilon_{yy} - \epsilon_{xx} \right) + \lambda_{E'} \left( \epsilon_{xz} + \epsilon_{zx} \right) \right] \left[ |E_x\rangle\langle E_x| - |E_y\rangle\langle E_y| \right] +$$

$$+ \left[ \lambda_E \left( \epsilon_{xy} + \epsilon_{yx} \right) + \lambda_{E'} \left( \epsilon_{yz} + \epsilon_{zy} \right) \right] \left[ |E_x\rangle\langle E_y| + |E_y\rangle\langle E_x| \right] \quad (6)$$

The orbital strain coupling constants $\lambda_{A_1}$, $\lambda_{A_1'}$, $\lambda_E$, and $\lambda_{E'}$ are in units of Hz.

In our experiments, the strain induced by the cantilever motion is linear in the cantilever tip deflection, $\epsilon = \epsilon_0 (a + a^\dagger)$. Here, $\epsilon_0$ is the strain induced by the cantilever zero point motion and $a$ is the annihilation operator of the cantilever mode. The single phonon coupling parameters, $g_\Gamma$, quantify the coupling of the orbital levels of the NV center to the strain of symmetry $\Gamma$ induced by the zero point motion of the cantilever.

### APPENDIX C: RESONANT EXCITATION SPECTROSCOPY (RES)

In Fig. 6b, we show the experimental sequence used for resonant excitation spectroscopy. The experiment begins with a short 1 μs pulse of 532 nm laser light generated by an acousto-optic modulator (AOM). This pulse initializes the NV into its negative charge state and also initializes it into the $m_s = 0$ spin level of the ground state, $|A\rangle$. Next, a 5 μs pulse of light near 637 nm from the tunable diode laser is sent to the NV for resonant excitation of the zero-phonon line (ZPL). During this pulse, we monitor the NV fluorescence using an avalanche photodiode. This two-pulse sequence is repeated 20,000 times for a particular laser frequency. The laser frequency is swept by changing the voltage of a piezoelectric element that controls the laser cavity length. In each scan, the laser is swept over a 38 GHz frequency band, which is the maximum frequency window allowed by the laser piezo. The laser frequency sweep is calibrated with a scanning Fabry-Perot invar cavity with a free spectral range of 1.5 GHz.

### APPENDIX D: RES SIGNAL UNDER AC STRAIN

When the cantilever is at rest, the RES signal contains two Lorentzian peaks corresponding to the $E_x$ and $E_y$ transitions, with each peak centered at $f_+$ and $f_-$ respectively. We typically observe optical linewidths of $\Gamma = 1$ GHz. When the cantilever vibrates, it introduces an AC strain field to the NV center, which modulates the optical transition frequencies, and hence the center frequency of these Lorentzian peaks. As shown in a later section, strain produced by the cantilever motion can also modulate polarization selection rules. For all RES experiments, the laser polarization is kept constant throughout the measurement, and hence the modulation of the polarization selection rules will manifest itself as a modulation of the PL emission intensity. Since our measurement time greatly exceeds the period of a single cantilever oscillation, our signal time averages over the cantilever's motion. Therefore, the signal can be described by

$$\mathcal{L}_\pm (f) = \frac{2\pi}{\omega_c} \int_0^{2\pi/\omega_c} \mathcal{I}_\pm (x(t)) \frac{(\Gamma/2)^2}{(\Gamma/2)^2 + (f - f_\pm(x(t)))^2} dt \qquad (7)$$

where $\mathcal{I}_\pm$ refers to the photoluminescence intensity of each peak (measured here in kC/s)

In addition, due to the finite quality factor of the cantilever, the cantilever will respond to mechanical excitations over a small frequency band around its resonance frequency. For weak mechanical excitation, the cantilever approximately exhibits a Lorentzian mechanical response whose linewidth is determined by its quality factor. Therefore, the normalized cantilever tip position $x(t)$ can be given by

$$x(t) = \frac{x_c(\omega_{piezo})}{x_0} \cos(\omega_{piezo} t) \quad (8)$$

where

$$x_c(\omega_{piezo}) \sim x_{max} \frac{(\gamma/2)^2}{(\gamma/2)^2 + (\omega_{piezo} - \omega_c)^2} \quad (9)$$

Here, $x_{max}$ refers to the maximum tip deflection when driven on resonance and $\gamma = \omega_c / Q$ is the linewidth of the mechanical resonance. The theoretical density plots shown in Fig. 2c. and Fig. 2e are calculated with the above equations, and incorporate experimental determinations of the intrinsic crystal strain, mechanical quality factor, optical linewidths, and polarization selection rules.

## APPENDIX E: STROBOSCOPIC MEASUREMENT PROCEDURE

The fundamental mechanical mode of the cantilever is actuated by a piezoelectric transducer that is electrically driven with a lock-in amplifier. During the measurement, both the frequency detuning from resonance and amplitude of the driven motion are kept constant with a phase-locked loop and feedback circuits. The amplitude of driven motion is measured with an interferometer whose signal is calibrated by the Brownian motion of the cantilever.

The stroboscopic measurement is accomplished by gating the photon detection. A home-built comparator circuit converts the mechanical drive signal from the lock-in amplifier into a TTL pulse train of the same frequency with a finite phase offset. The duty cycle and hence the width of the "ON" region of the pulse train is controlled by the threshold voltage of the comparator chip. The finite phase offset is tuned with a phase shifter, and can be tuned over a full 360 degrees. This pulse train directly gates the APD through an external modulation input. For the data shown in Fig. 3b-d, the TTL pulse train is frequency doubled, allowing for measurements of two cantilever positions that are 180 degrees out of phase from each other.

## APPENDIX F: STROBOSCOPIC RES SIGNAL

During stroboscopic measurements, RES signals average over a small segment of the cantilever motion, as described in the Methods section. Therefore, the general stroboscopic RES signal can be written as

$$\mathcal{L}_{\pm}^{strob}(f) = \frac{1}{\tau} \int_{T}^{T+\tau} \mathcal{I}_{\pm}(x(t)) \frac{\left(\Gamma/2\right)^2}{\left(\Gamma/2\right)^2 + \left(f - f_{\pm}(x(t))\right)^2} dt \qquad (10)$$

Where $T$ indicates the start time of the PL detection and $\tau$ is the length of the detection window of 60 ns. The start time, $T$, is set by the relative phase between the TTL pulse train gating the single photon counter and the mechanical driving signal. In our experiments, we set $T$ such that the detection window lies symmetrically around an antinode of the cantilever's motion. For instance, when we are interested in the maximum downward deflection of the cantilever (corresponding to negative $x$), we set $T = \frac{\pi}{\omega_c} - \frac{\tau}{2}$.

The detection window is much shorter than the oscillation period of the cantilever, and we approximately measure the NV spectral response for a well-defined positon of the cantilever. When the detection window is placed symmetrically around an antinode of motion, the stroboscopic RES signal can be approximated by a Lorentzian.

$$\mathcal{L}_{\pm}^{strob}(f) \approx \mathcal{I}_{\pm}(x_{max/min}) \frac{\left(\Gamma/2\right)^2}{\left(\Gamma/2\right)^2 + \left(f - f_{\pm}(x_{max/min})\right)^2} \qquad (11)$$

Here, $x_{max/min}$ is the maximum or minimum normalized displacement of the cantilever tip depending on which antinode is selected. The fits to the stroboscopic RES data shown in Fig. 3a in the main text reflect this approximation. For data shown in Figs. 3b-d and in Figs. 5f-h, the center frequency and amplitudes of the stroboscopic RES peaks as calculated by fits to eq. (9) are used respectively.

## APPENDIX G: STRAIN PROFILE OF A SINGLY-CLAMPED CANTILEVER

The mechanical modes of our diamond cantilevers are well-described by Euler-Bernoulli beam theory. We consider a singly-clamped diamond beam of length $l$, width $w$, and thickness $t$. The cantilever coordinate system $(X, Y, Z)$ can be described in terms of the diamond crystallographic directions, as shown in Fig. 7a. The $Z$-axis is the cantilever axis, and is parallel to the [110] crystal direction. The strain profile for the cantilever's fundamental flexural motion along the $Y$ direction as function of axial position $Z$ and cantilever tip deflection $x_c$ is given by

$$\varepsilon(Z, x_c) \simeq R_0 \frac{x_c}{2l^2}(1.875)^2 \left[ \cos\left(1.875\frac{Z}{l}\right) + \cosh\left(1.875\frac{Z}{l}\right) - \frac{1}{1.3622}\left(\sin\left(1.875\frac{Z}{l}\right) + \sinh\left(1.875\frac{Z}{l}\right)\right) \right]$$

(12)

where $R_0$ refers to the distance in the $Y$ direction from the center of the beam. For a more detailed derivation, we refer the reader to ref. [17].

For an NV that is at a depth $d_i$ from the surface of the cantilever, $R_0 = t/2 - d_i$. For the devices in this paper, the NV depth is determined by the nitrogen implantation process, and $d_i = 51.5 \pm 13.0$ nm, where the uncertainty in $d_i$ is derived from Stopping Range of Ions in Matter (SRIM) simulations, representing one standard deviation in the straggle of nitrogen ions during ion implantation.

## APPENDIX H: STRAIN TENSORS IN THE NV BASIS

The strain tensor generated by the cantilever's flexural motion can be written in the cantilever's coordinate system $(X, Y, Z)$ as follows

$$\epsilon_c = \begin{pmatrix} -\nu\varepsilon & 0 & 0 \\ 0 & -\nu\varepsilon & 0 \\ 0 & 0 & \varepsilon \end{pmatrix} \qquad (13)$$

where $\varepsilon$ is the mechanically induced strain along the $Z$ direction as defined in eq. (12) and $\nu = 0.11$ is the Poisson ratio of diamond [44]. Our cantilevers are fabricated such that $X \| [\bar{1}10]$, $Y \| [001]$ and $Z \| [110]$.

We transform the above strain tensor from the cantilever coordinate system, $(X, Y, Z)$, into the coordinate system of the NV center, $(x, y, z)$. For NVs in group A ($z \| [\bar{1}\bar{1}\bar{1}]$ or $[11\bar{1}]$), the transformed strain tensor is

$$\epsilon_A = \begin{pmatrix} \frac{\varepsilon}{3}(1-2\nu) & 0 & -\frac{\sqrt{2}\varepsilon}{3}(1+\nu) \\ 0 & -\nu\varepsilon & 0 \\ -\frac{\sqrt{2}\varepsilon}{3}(1+\nu) & 0 & \frac{\varepsilon}{3}(2-\nu) \end{pmatrix} \qquad (14)$$

For NVs in group B ($z \parallel [\bar{1}11]$ or $[1\bar{1}1]$), the transformed strain tensor is

$$\epsilon_B = \begin{pmatrix} -\nu\varepsilon & 0 & 0 \\ 0 & \varepsilon & 0 \\ 0 & 0 & -\nu\varepsilon \end{pmatrix} \tag{15}$$

With the above strain tensors, we can diagonalize the Hamiltonian in eq. (6) and write the $E_x$ and $E_y$ transition frequencies for NVs in both groups in terms of $\varepsilon$. We obtain

$$f_\pm^A = f_{ZPL} + \delta f_{A_1} + \lambda_{A_1}\frac{2-\nu}{3}\varepsilon + \lambda_{A_1'}\frac{1-5\nu}{3}\varepsilon \pm \sqrt{\left(-\lambda_E\frac{1+\nu}{3}\varepsilon - \lambda_{E'}\frac{2\sqrt{2}(1+\nu)}{3}\varepsilon + \delta f_{E_1}\right)^2 + \left(\delta f_{E_2}\right)^2}$$

for group A and

$$f_\pm^B = f_{ZPL} + \delta f_{A_1} - \lambda_{A_1}\nu\varepsilon + \lambda_{A_1'}(1-\nu)\varepsilon \pm \sqrt{\left(\lambda_{E_1}(1+\nu)\varepsilon + \delta f_{E_1}\right)^2 + \left(\delta f_{E_2}\right)^2} \text{ for group B.}$$

## APPENDIX I: CHARACTERIZATION OF INTRNSIC STRAIN AND POLARIZATION SELECTION RULES WITH STRAIN

To properly model the NV response to crystal strain and extract the single phonon coupling parameters, it is necessary to characterize the intrinsic local $E$ symmetric strain environment of the NV center. The intrinsic strain can be completely characterized by measuring the splitting between the $|E_x\rangle$ and $|E_y\rangle$ levels and measuring the orientation of the transition dipole moments. For intrinsic strains $\delta f_{E_1}$ and $\delta f_{E_2}$, we see that the splitting between the $|E_x\rangle$ and $|E_y\rangle$ levels can be written as

$$\Delta f_0 = 2\sqrt{(\delta f_{E_1})^2 + (\delta f_{E_2})^2} \tag{16}$$

The orientation of the transition dipole moments can be extracted by measuring the polarization dependence of the $E_x$ and $E_y$ transitions. The interaction of the resonant laser with the NV is described by electric dipole interaction, $H_E = -\vec{d}\cdot\vec{E}$ where $\vec{d}$ is the NV electric dipole moment and $\vec{E}$ is the electric field of the excitation laser. The axial dipole moment, $d_z$, does not connect the ground and excited state manifolds, but merely shifts them relative to each other. The transverse electric dipole moments $d_x$ and $d_y$ split and mix the $|E_x\rangle$ and $|E_y\rangle$ levels, but also couple the ground and excited state manifolds.

In the $\{|A\rangle, |E_x\rangle, |E_y\rangle\}$ basis and in the absence of strain, the transverse electric dipole moments $d_x$ and $d_y$ can be expressed as

$$d_x = \begin{pmatrix} 0 & 0 & d_\perp \\ 0 & d_\perp & 0 \\ d_\perp & 0 & -d_\perp \end{pmatrix} \tag{17}$$

$$d_y = \begin{pmatrix} 0 & d_\perp & 0 \\ d_\perp & 0 & d_\perp \\ 0 & d_\perp & 0 \end{pmatrix} \tag{18}$$

Clearly, the $E_x$ and $E_y$ transitions are driven by linearly polarized light along the $\hat{x}$ and $\hat{y}$ directions of the NV center. In the presence of $E$-symmetric strain, $|E_x\rangle$ and $|E_y\rangle$ are mixed, and new eigenstates are formed. Simultaneously, this leads to a rotation of the transverse electric dipole moments. The rotated dipole operators, $d_x$ and $d_y$ can be expressed as

$$d_x = \cos(\theta) d_x + \sin(\theta) d_y \tag{19}$$

$$d_y = -\sin(\theta) d_x + \cos(\theta) d_y \tag{20}$$

where $2\theta$ is the Stuckelberg angle and $\tan(2\theta) = \dfrac{(\delta f_{E_2} + g_{E_2} x)}{(\delta f_{E_1} + g_{E_1} x)}$. Based on the crystal orientations and the polarization angle, $\phi$ defined in Fig. 5a, the electric field of the incident light ($\vec{k} \parallel [00\bar{1}]$) can be written as $\vec{E} = |E| \left( \dfrac{\cos(\phi)}{\sqrt{2}} [\bar{1}10] + \dfrac{\sin(\phi)}{\sqrt{2}} [110] \right)$. The absorption intensity is proportional to $|\vec{d} \cdot \vec{E}|^2$. Therefore, the normalized absorption intensity of the two excited states are given by

$$I_{E_x} = \left( \frac{\cos(\theta)\cos(\phi)}{\sqrt{3}} - \sin(\theta)\sin(\phi) \right)^2, \quad I_{E_y} = \left( \frac{\sin(\theta)\cos(\phi)}{\sqrt{3}} + \cos(\theta)\sin(\phi) \right)^2 \text{ for group A and}$$

$$I_{E_x} = \left( \frac{\cos(\theta)\sin(\phi)}{\sqrt{3}} - \sin(\theta)\cos(\phi) \right)^2, \quad I_{E_y} = \left( \frac{\sin(\theta)\sin(\phi)}{\sqrt{3}} + \cos(\theta)\cos(\phi) \right)^2 \text{ for group B.}$$

In our polarization dependence measurements, we need to account for saturation of the optical transitions and non-zero ellipticity of the excitation laser. Here, we assume a simple model of saturation in which the normalized absorption intensity is given by $I = 1 - e^{-P_{\text{eff}}/P_{\text{sat}}}$, where $P_{\text{eff}}$ is the effective laser power seen by the NV and $P_{\text{sat}}$ is the saturation power. The effective laser power, $P_{\text{eff}}$, is determined by the polarization dependence of the transition as well as the total incident laser power, $P_{\text{in}}$. To account for ellipticity, we introduce a relative phase delay, $\psi$, between the x and y components of the electric field: $\vec{E} = |E|\left(\frac{\cos(\phi)}{\sqrt{2}}[\bar{1}10] + e^{i\psi}\frac{\sin(\phi)}{\sqrt{2}}[110]\right)$.

Putting it all together, the normalized absorption intensities for the $E_x$ and $E_y$ transitions including these two effects are given by

$$I_{E_x} = 1 - \exp\left[-\frac{P_{in}}{P_{sat}}\left(\sin^2(\theta)\sin^2(\phi) + \frac{\cos^2(\theta)\cos^2(\phi)}{3} - \frac{\cos(\psi)\sin(2\theta)\sin(2\phi)}{2\sqrt{3}}\right)\right],$$

$$I_{E_y} = 1 - \exp\left[-\frac{P_{in}}{P_{sat}}\left(\cos^2(\theta)\sin^2(\phi) + \frac{\sin^2(\theta)\cos^2(\phi)}{3} + \frac{\cos(\psi)\sin(2\theta)\sin(2\phi)}{2\sqrt{3}}\right)\right] \text{ for group A}$$

and

$$I_{E_x} = 1 - \exp\left[-\frac{P_{in}}{P_{sat}}\left(\sin^2(\theta)\cos^2(\phi) + \frac{\cos^2(\theta)\sin^2(\phi)}{3} - \frac{\cos(\psi)\sin(2\theta)\sin(2\phi)}{2\sqrt{3}}\right)\right],$$

$$I_{E_y} = 1 - \exp\left[-\frac{P_{in}}{P_{sat}}\left(\cos^2(\theta)\cos^2(\phi) + \frac{\sin^2(\theta)\sin^2(\phi)}{3} + \frac{\cos(\psi)\sin(2\theta)\sin(2\phi)}{2\sqrt{3}}\right)\right] \text{ for group B.}$$

The fits of the polarization plots are shown in Fig. 5 as well as Figs. 8 and 9. where $\theta$, $P_{sat}$ and $\psi$ are used as fit parameters. The fit results obtained from 12 NVs yield $P_{sat} = 0.4 \pm 0.1\,\mu W$, $\psi = 54 \pm 10°$ and the results of $\theta$ are used to determine the intrinsic strain.

## APPENDIX J: EXTRACTING THE STRAIN COUPING CONSTANTS AND ERROR BARS

To measure the strain-orbit coupling constants, we measure the $E_x$ and $E_y$ transition frequencies, $f_+$ and $f_-$, for NVs in groups A and B as function of the strain induced along the cantilever axis. Flexural motion of the cantilever introduces a strain $\varepsilon$ along the cantilever axis, while producing strains $-\nu\varepsilon$ in the transverse directions due to the Poisson effect ($\nu = 0.11$). Under the cantilever strain, the $E_x$ and $E_y$ transition frequencies can be written as

$$f_\pm^A(\varepsilon) = f_{ZPL} + \delta f_{A_1} + \lambda_{A_1}\frac{2-\nu}{3}\varepsilon + \lambda_{A_1'}\frac{1-5\nu}{3}\varepsilon$$

$$\pm\sqrt{\left(-\lambda_E\frac{1+\nu}{3}\varepsilon - \lambda_{E'}\frac{2\sqrt{2}(1+\nu)}{3}\varepsilon + \delta f_{E_1}\right)^2 + \left(\delta f_{E_2}\right)^2} \quad (21)$$

$$f_\pm^B(\varepsilon) = f_{ZPL} + \delta f_{A_1} - \lambda_{A_1}\nu\varepsilon + \lambda_{A_1'}(1-\nu)\varepsilon \pm \sqrt{\left(\lambda_E(1+\nu)\varepsilon + \delta f_{E_1}\right)^2 + \left(\delta f_{E_2}\right)^2} \qquad (22)$$

To extract the $A_1$-symmetric strain coupling constants, we monitor the uniform shifts of the $E_x$ and $E_y$ transition frequencies as a function of $\varepsilon$, $\Delta_{A_1} = \dfrac{f_+ + f_-}{2}$. Neglecting the static frequency offset, $f_{ZPL} + \delta f_{A_1}$, we obtain

$$\Delta_{A_1}{}^A(\varepsilon) = \lambda_{A_1}\frac{2-\nu}{3}\varepsilon + \lambda_{A_1'}\frac{1-5\nu}{3}\varepsilon \qquad (23)$$

$$\Delta_{A_1}{}^B(\varepsilon) = -\lambda_{A_1}\nu\varepsilon + \lambda_{A_1'}(1-\nu)\varepsilon \qquad (24)$$

The vertical error bars in Fig. 3b-d reflect the uncertainty in peak positions in the RES measurements, which is dominated by the 5° phase uncertainty in the location of the stroboscopic measurement window with respect to the cantilever's motion. The horizontal error bars in Fig. 3b-d are given by the uncertainty in 3-D position of the NV, which is dominated by the 13 nm uncertainty in the NV depth in the diamond, which is the expected straggle of nitrogen ions during our implantation process (as calculated by SRIM). To extract $\lambda_{A_1}$ and $\lambda_{A_1'}$, we simultaneously fit the data in Fig. 3b to the above equations for $\Delta_{A_1}$. The error reported for $\lambda_{A_1}$ and $\lambda_{A_1'}$ is limited by our calibration in the uncertainty of $\varepsilon$, which is determined in the uncertainty in the optical interferometry measurement of the cantilever tip deflection. We estimate a 15% uncertainty in this calibration, and hence quote a 15% uncertainty in the values of $\lambda_{A_1}$ and $\lambda_{A_1'}$.

With this information, we next extract $\lambda_E$ and $\lambda_{E'}$ by extracting the overall shifts of the $E_x$ and $E_y$ transition frequencies for 12 NVs (6 NVs in each group), as a function of $\varepsilon$. The uncertainty in the measured values of $\lambda_E$ and $\lambda_{E'}$ also reflect the 15% uncertainty in the interferometric measurement of the cantilever tip deflection. Data for all measured NVs are shown in Figs. 10 and 11.

### APPENDIX K: COOPERATIVITY AND RESONATOR COOLING CALCULATIONS

The device described in the main text is a doubly-clamped diamond nanobeam with dimensions (2 μm, 100 nm, 50 nm) containing a near-surface NV center located at one of the clamping points. We will focus on the fundamental mode of this resonator, which has a resonance frequency $\omega_c/2\pi = 238\,\text{MHz}$. Moreover, we assume that the nanobeam has the same crystallographic orientation as the device shown in this paper.

In this calculation, we will consider an NV center from group B, whose symmetry axis is transverse to the resonator axis. For simplicity, we will assume that the NV is hosted in a perfect crystal with no intrinsic strain. Using the strain tensor in eq. (15), the NV-resonator interaction Hamiltonian is

$$H_{int} = \left[-\lambda_{A_1}\nu\varepsilon + \lambda_{A_1'}(1-\nu)\varepsilon\right]\left[|E_x\rangle\langle E_x| + |E_y\rangle\langle E_y|\right] + \left[\lambda_E(1+\nu)\varepsilon\right]\left[|E_x\rangle\langle E_x| - |E_y\rangle\langle E_y|\right] \quad (25)$$

where $\varepsilon$ is the strain induced along the axis of the nanobeam. Next, we quantize the strain field in terms of the phonon creation and annihilation operators for the resonator mode such that $\varepsilon = \varepsilon_0(a+a^\dagger)$ and $\varepsilon_0$ is the strain due to zero point motion. Based on the signs of the strain coupling constants, the highest single phonon coupling occurs for the $A \rightarrow E_y$ transition. We can isolate this transition and neglect $|E_x\rangle$ by using $\hat{y}$ polarized light. We find the effective interaction Hamiltonian for the $|E_y\rangle$ state is

$$H_{eff} = g_\parallel |E_y\rangle\langle E_y|(a+a^\dagger) \quad (26)$$

where $g_\parallel = \left[-\lambda_{A_1}\nu + \lambda_{A_1'}(1-\nu) - \lambda_E(1+\nu)\right]\varepsilon_0 = (2.31\,\text{PHz})\varepsilon_0$ is the effective single-phonon coupling parameter. For the device described here, we expect $g_\parallel = 21.5\,\text{MHz}$ for the fundamental mode.

The single phonon cooperativity is defined to be $\eta = g_\parallel^2 / \Gamma_2 \gamma_{th}$, where $\Gamma_2$ is the dephasing rate of the optical transition, and $\gamma_{th}$ is the heating rate of the mechanical resonator [10]. The heating rate, $\gamma_{th}$, is defined by $\gamma_{th} = \bar{n}\omega_c / 2\pi Q$, where $\bar{n}$ is the equilibrium phonon occupation number of the resonator at temperature $T$, and $Q$ is the quality factor of the mechanical mode. For the calculation of the cooperativity in the main text, we assume a value of $\Gamma_2 = 100\,\text{MHz}$ for the $A \rightarrow E_y$ transition, which has been observed in ref. [45]. This value is limited primarily by electron-phonon interactions within the excited state manifold. In addition, we assume that the quality factor of the mechanical mode is $Q = 10^5$, which has already been demonstrated with micron-scale diamond resonators [22,46]. At a temperature $T = 4\,\text{K}$, the diamond resonator has an equilibrium thermal occupation number of $\bar{n} = 367$. Putting it all together, we find the cooperativity of this device to be $\eta = 5.2$.

To evaluate the device's prospects for ground state cooling, we consider the protocol for sideband cooling as detailed in refs. [13,14]. In this protocol, we resonantly address the first red vibrational sideband of the $A \rightarrow E_y$ transition associated with the fundamental mechanical mode of the resonator. The steady state thermal occupation number for the resonator under sideband cooling obeys

$$\bar{n} \approx \frac{\gamma_{th}}{\Gamma_C} \qquad (27)$$

where $\Gamma_C$ is the sideband cooling rate. The sideband cooling rate is given by

$$\Gamma_C = 4\pi^2 \frac{g_\parallel^2 \Omega^2}{\Gamma \omega_c^2} \qquad (28)$$

where $\Omega$ is the optical Rabi frequency and $\Gamma$ is the linewidth of the optical transition. Assuming an optical Rabi frequency $\Omega = 100$ MHz, a linewidth of $\Gamma = 100$ MHz, and the $\gamma_{th}$ described above, we find $\Gamma_C = 843$ kHz and $\bar{n} \approx 0.9$.

**Figure captions**

Figure 1: **A strain-coupled NV-mechanical resonator device a,** An NV center is embedded in a diamond cantilever and strain produced by the cantilever motion shifts the resonance of the zero-phonon line. A 637 nm laser probes the resonant shifts using resonant excitation spectroscopy. **b,** A photoluminescence image of a diamond cantilever shows the presence of embedded NV centers **c,** Simplified energy level diagram depicting the optical transitions between the $m_s = 0$ electronic levels. Under no strain, the $E_x$ and $E_y$ states are degenerate (left). Strain of $A_1$ symmetry shifts the $E_x$ and $E_y$ states together (center) while strain of $E$ symmetry splits the $E_x$ and $E_y$ levels (right).

Figure 2: **Hybrid defect-resonator coupling. a,** RES measurement of a single NV center embedded in a cantilever. In the absence of mechanical driving (light gray), the optical transition lineshapes are Lorentzian. Under resonant mechanical excitation (green), the transitions are modulated and the lineshapes are broadened. **b,** Experimental and **c,** simulated excitation spectra of the $E_x$ and $E_y$ optical transitions as a function of mechanical drive detuning, showing the resonant nature of the coupling. **d,** Experimental and **e,** simulated excitation spectra of the $E_x$ and $E_y$ transitions as a function of the amplitude of resonant mechanical excitation.

Figure 3: **Quantifying the strain coupling constants. a**, A stroboscopic RES measurement shows shifts of the $E_x$ and $E_y$ transitions when the cantilever tip is deflected upward (blue) or downward (yellow) by 24 nm. The transitions when the cantilever is at rest are shown in gray. The lineshapes are fit with Lorentzians (black) **b,** Plot of the common mode shift of the $E_x$ and $E_y$ states for 12 different NVs, each marked by different colors. Circles (triangles) indicate NVs in group A (B). Linear fits to the data are shown in black. The gray shaded area indicates a 15% uncertainty in the amplitude of driven motion determined from optical interferometry measurements. **c (d)** plot the total frequency shifts of the $E_x$ and $E_y$ transitions for NVs in group A(B). Error bars in (**b-d**) correspond to $5°$ phase uncertainties in the relative phase between the measurement window and the cantilever's motion in the stroboscopic measurement (vertical) and 13 nm uncertainties in the NV depth (horizontal).

Figure 4: **Strain-mediated frequency matching of two NVs. a-c**, RES measurements for two NVs (NV I, NV II). With no mechanical driving, the $E_x$ and $E_y$ transitions of NV I (**a**) differ in frequency from those of NV II (**c**) **b**, RES measurements of NV I for increasing cantilever tip deflection amplitudes. At amplitudes of 30 nm and 43 nm respectively, the $E_x$ and $E_y$ transition frequencies match those of NV II, as indicated by red circles.

Figure 5: **Strain-mediated polarization matching of NVs. a**, Schematic showing the electric field vector (red arrow) of the excitation laser ($\hat{k} \| [00\bar{1}]$) and the linear polarization angle, $\phi$, defined with respect to the $[\bar{1}10]$ crystal axis. **b-e**, $\phi$-dependent RES measurements for NVs A-D with no mechanical excitation showing the distinct polarization dependences of the $E_x$ and $E_y$ transitions due to local, intrinsic strain. Data points plot the PL amplitudes of the $E_x/E_y$ peaks versus $\phi$ and solid lines are fits (see

Appendix I). **f-h**, $\phi$-dependent stroboscopic RES measurements for NVD at tip deflections of (-9.4 nm, -3.7 nm, -12.4 nm), matching the dipole excitation pattern of NV D to those of NVs A-C. Dashed lines are theoretical predictions using the tip deflection amplitude as input.

Figure 6: **Schematic of the experimental setup and the optical pulse sequence. a**, A schematic of our home-built, cryogenic scanning confocal microscope and the relevant optical paths for the 532 nm, 637 nm, and 450 nm lasers. The diamond resonator is located inside of closed-cycle cryostat (T < 7 K) and its mechanical motion is driven by a piezoelectric actuator. **b**, Pulse sequence used for resonant excitation spectroscopy. A 532 nm laser pulse (1 μs) initializes the NV to $|A\rangle$ and stabilizes the charge state. Next, a ~637 nm laser pulse (5 μs) is used for excitation spectroscopy of the $E_x$ and $E_y$ transitions. Photons emitted into the NV phonon sideband are collected during this excitation pulse. We repeat this sequence for N ~ 20,000 at each laser detuning $\Delta$ over the range of ± 19 GHz.

Figure 7: **Cantilever geometry and NV structure. a**, Schematic showing the geometry and orientation of our diamond cantilevers. The cantilever coordinate system $(X, Y, Z)$ is defined in terms of the diamond crystallographic orientations. **b**, To the left, we show the NV structure and coordinate system $(x, y, z)$. Black atoms represent carbon atoms, the dotted white atom represents the vacancy, and the blue atom represents the nitrogen atom. The NV $z$-axis lies along its symmetry axis, which is parallel to the bond connecting the nitrogen atom to the vacancy. To the right, the NV is being viewed along its symmetry axis.

Figure 8: **Group A polarization dependence measurements for extraction of intrinsic strain. a-f**, Normalized PL intensity is plotted as a function of the 637 nm laser linear polarization angle for 6 different NVs in group A. The cantilever is at rest. $E_x$ ($E_y$) states are marked with red triangles (blue circles). Solid lines are fits as described below.

Figure 9: **Group B polarization dependence measurements for extraction of intrinsic strain. a-f**, Normalized PL intensity is plotted as a function of the 637 nm laser linear polarization angle for 6 different NVs in group B. The cantilever is at rest $E_x$ ($E_y$) states are marked with red triangles (blue circles). Solid lines are fits as described below.

Figure 10: **Shifts of the $E_x$ and $E_y$ states as a function of strain (group A). a-f**, The energy levels of the $E_x$ and $E_y$ states for group A NV centers are plotted as a function of mechanical strain induced along the cantilever axis. $E_x$ ($E_y$) states are marked with blue squares (orange triangles). We simultanesoulsy fit the data with $\lambda_E$ and vertical offsets (solid lines). The gray shaded area indicates 15 % uncertainty in the calibration of the cantilever amplitude.

Figure 11: **Shifts of the $E_x$ and $E_y$ states as a function of applied strain (group B). a-f**, The energy levels of the $E_x$ and $E_y$ states for group B NV centers are plotted as a function of mechanical strain induced along the cantilever axis. $E_x$ ($E_y$) states are marked with blue squares (orange triangles). We simultanesoulsy fit the data with $\lambda_{E'}$ and vertical offsets (solid lines). The gray shaded area indicates 15 % uncertainty in the calibration of the cantilever amplitude.

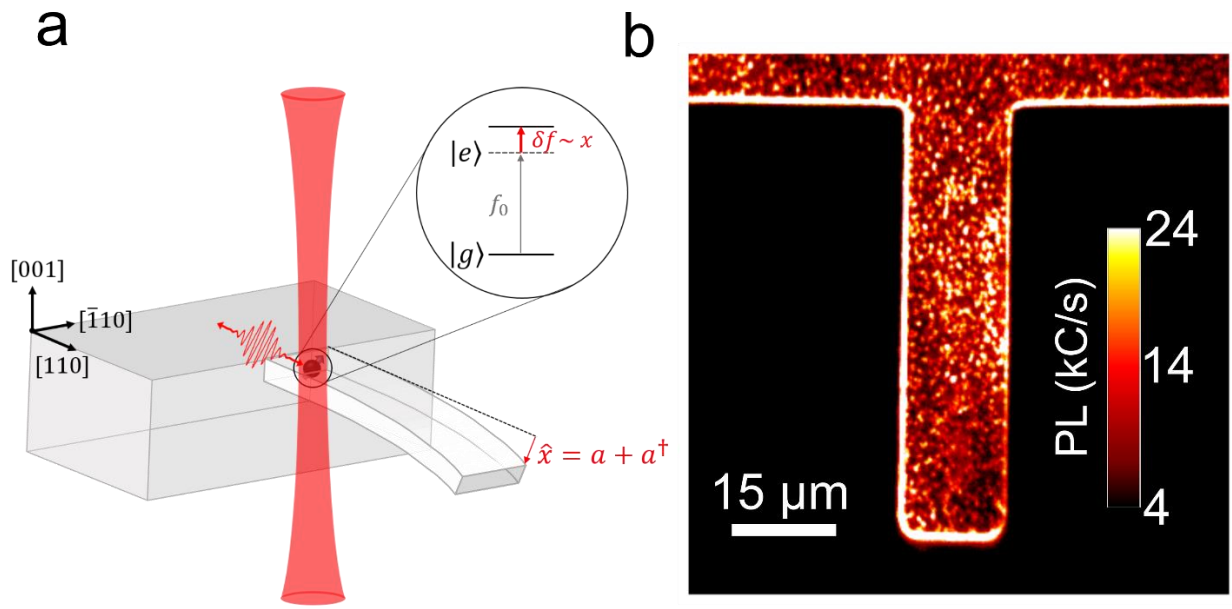

Fig. 1

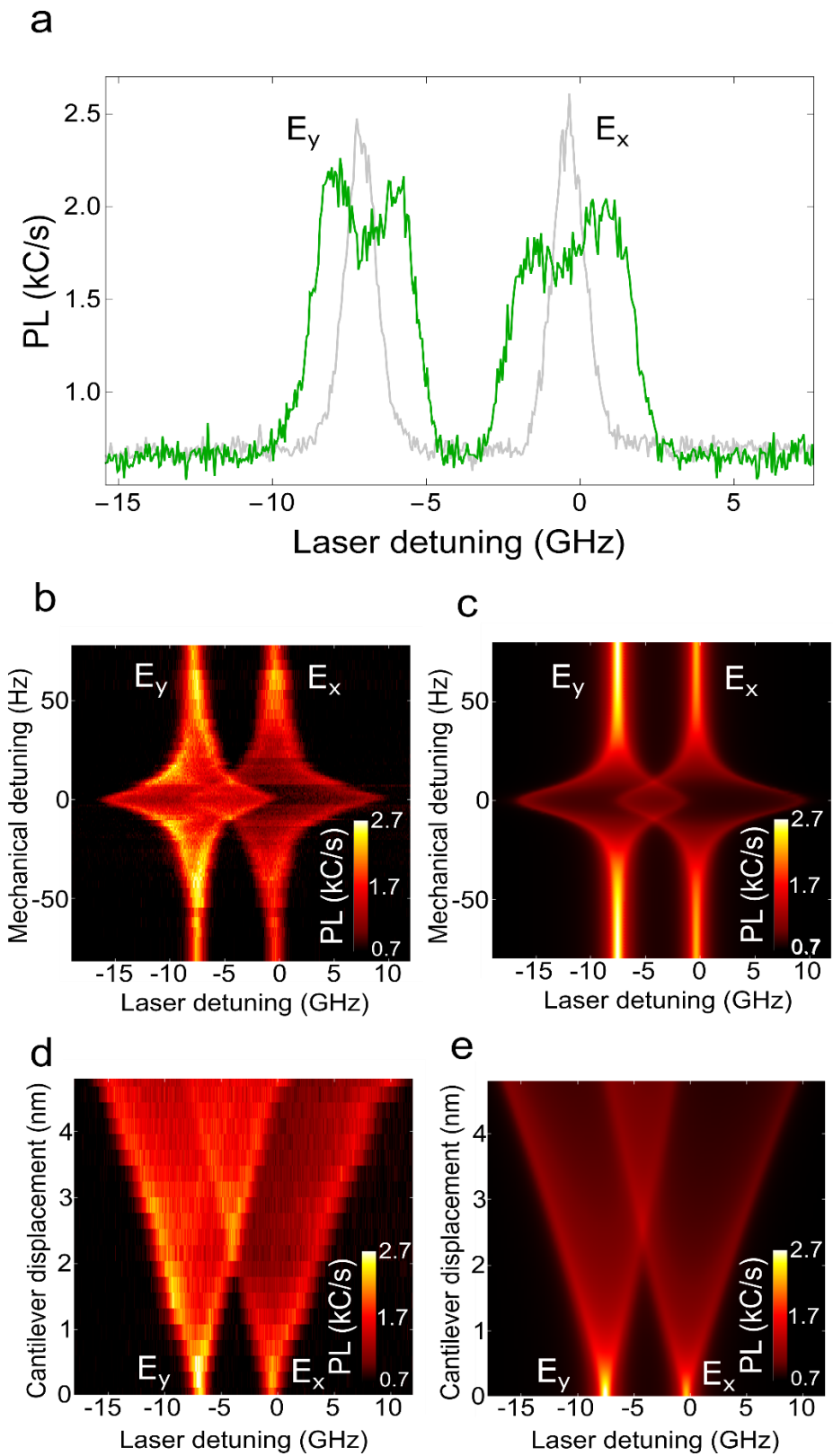

Fig. 2

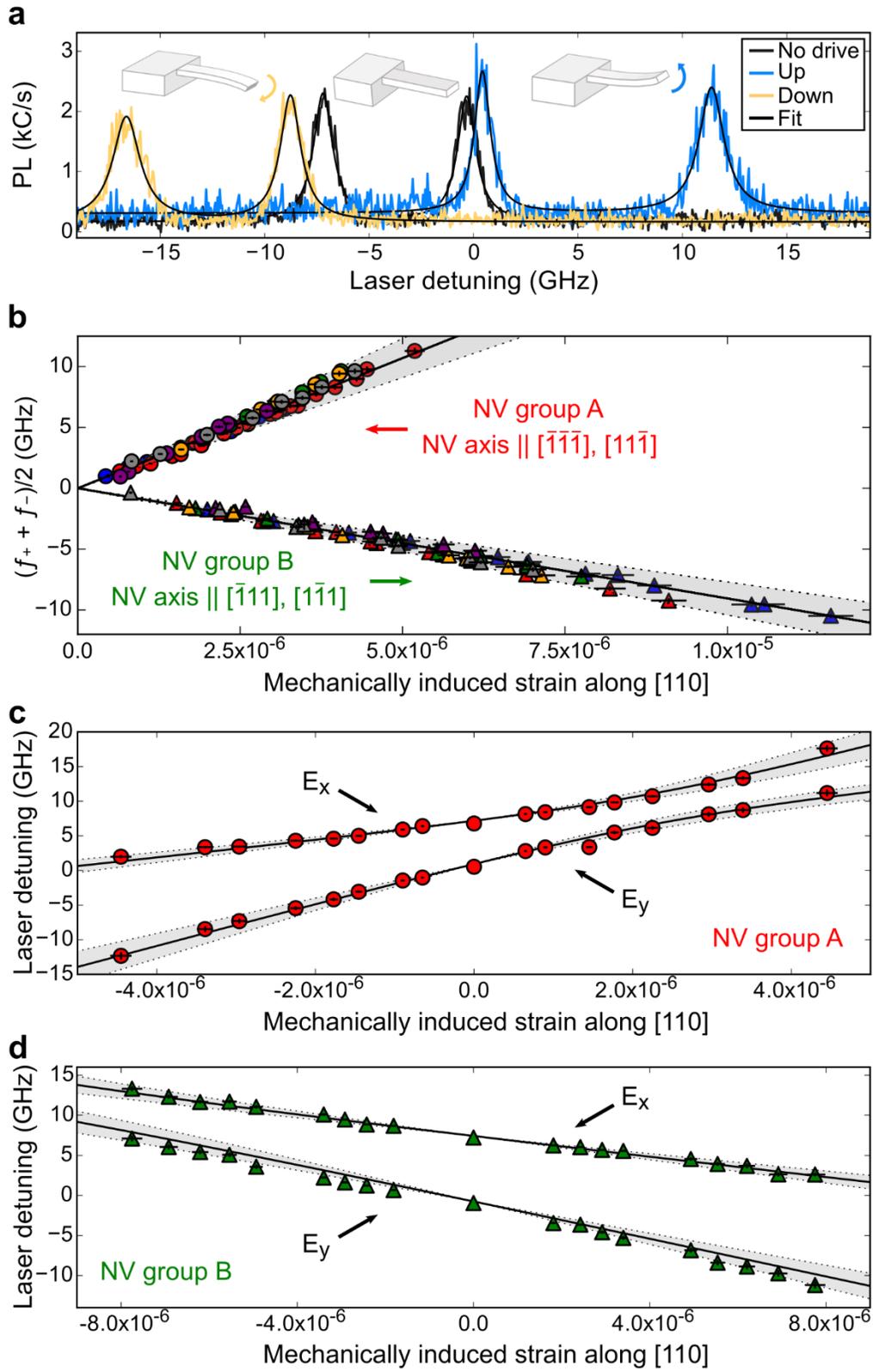

Fig. 3

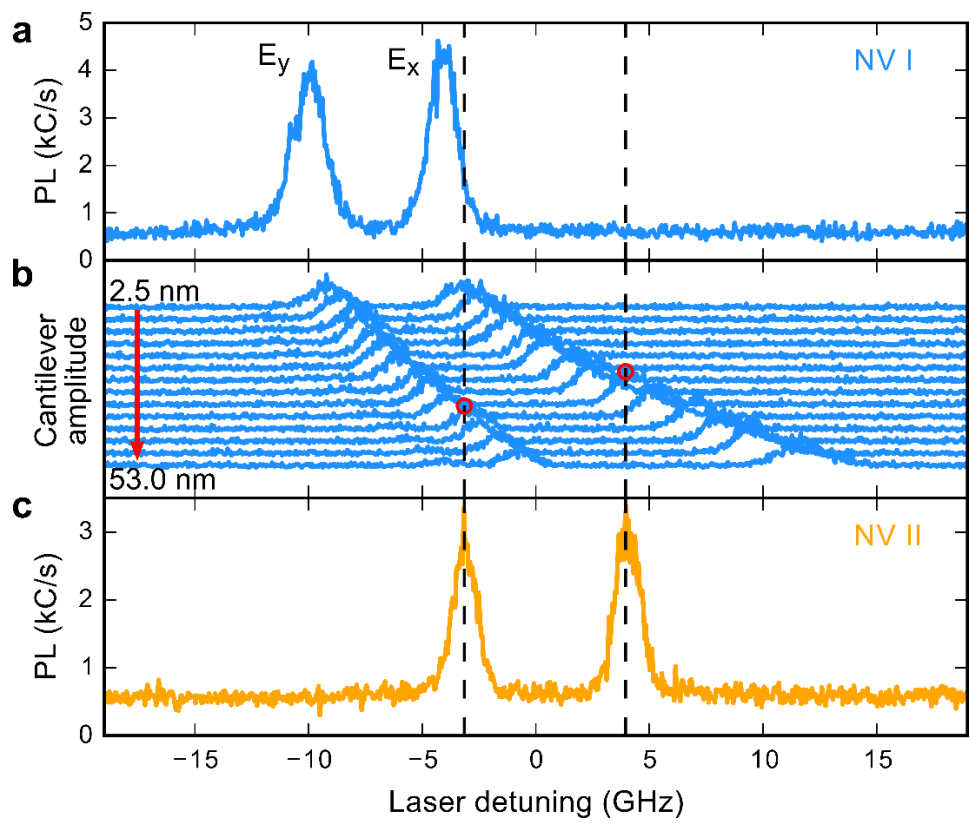

Fig. 4

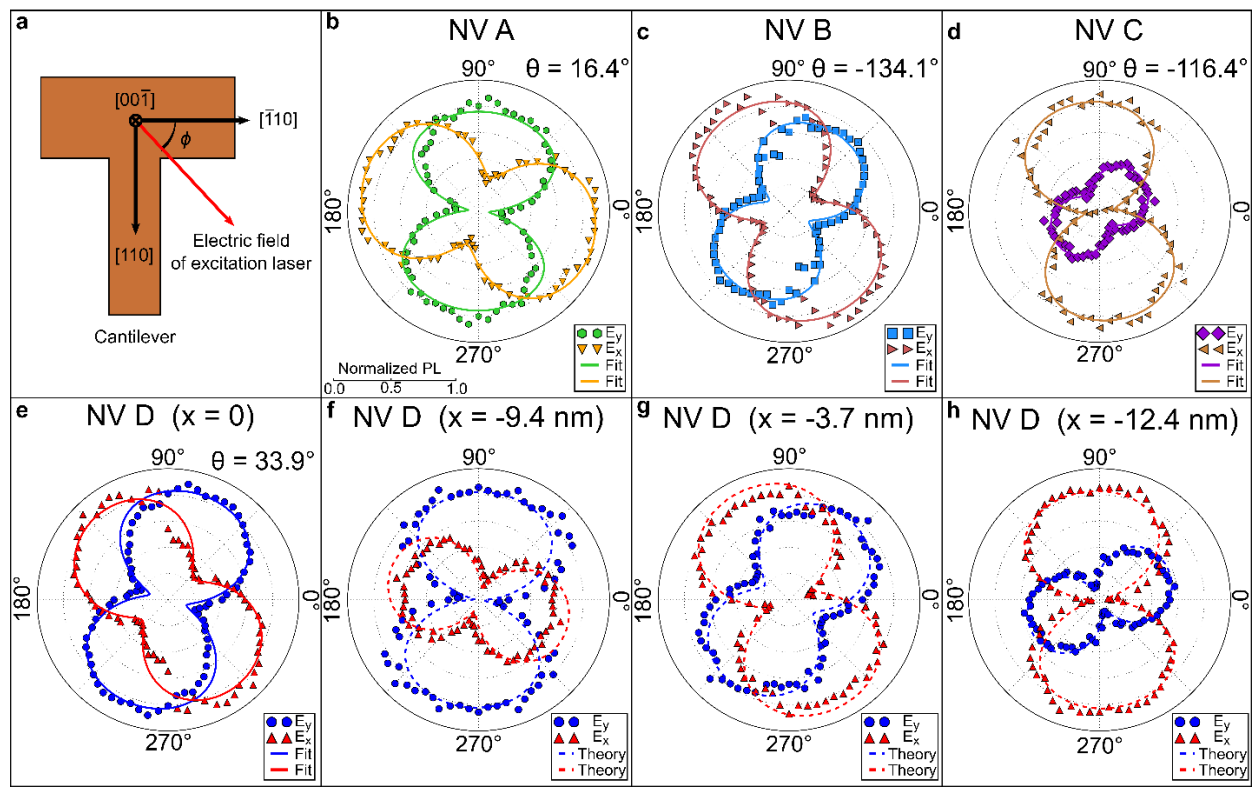

Fig. 5

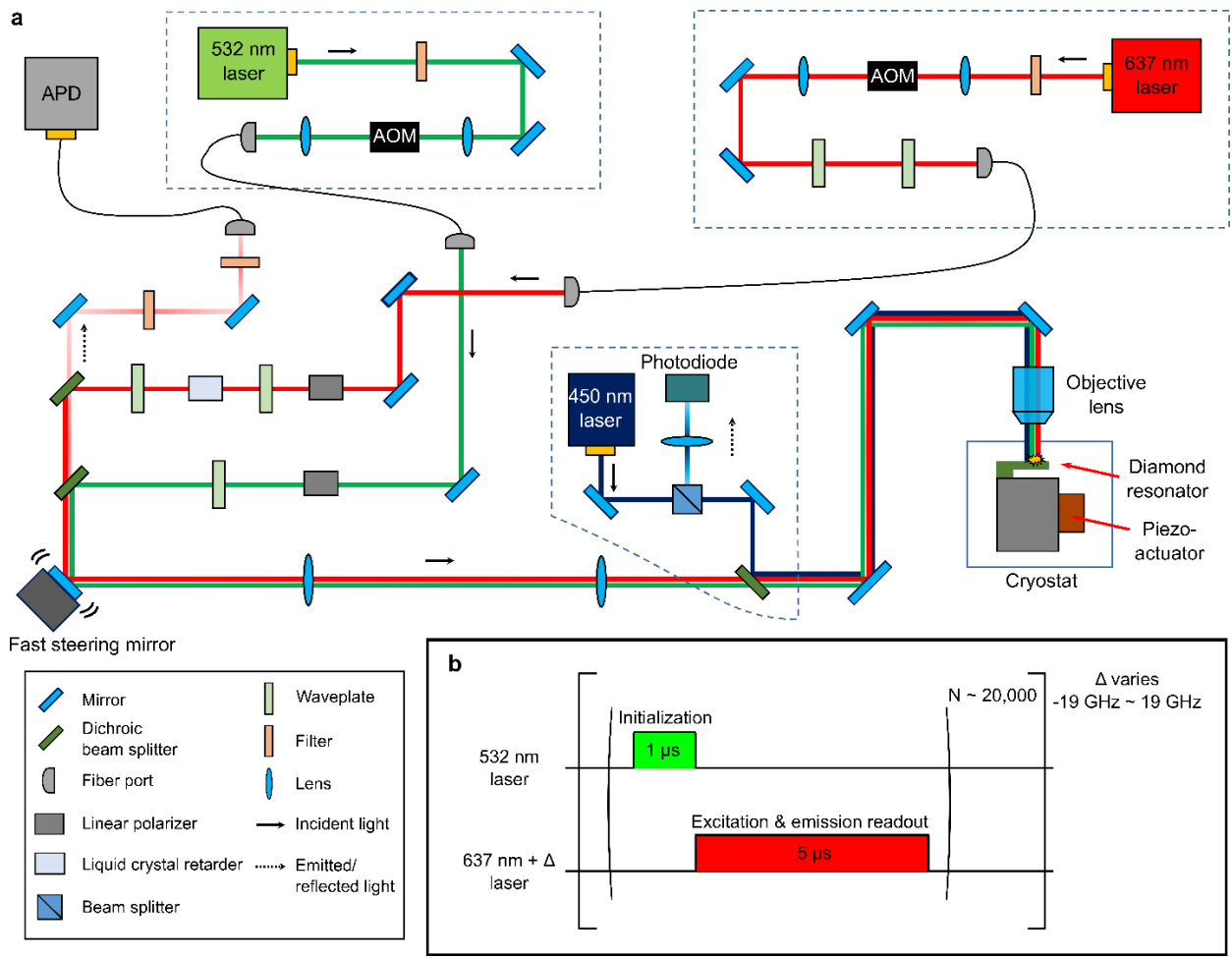

Fig. 6

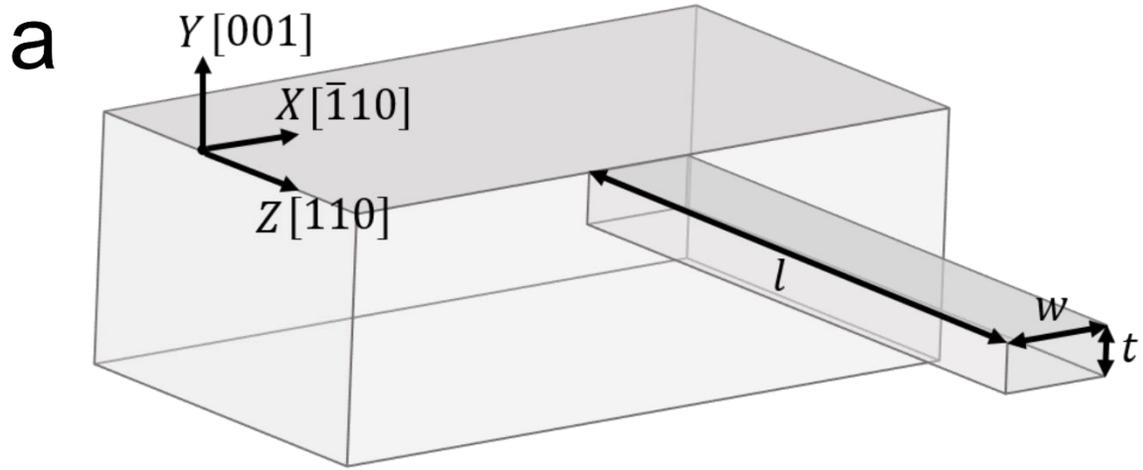

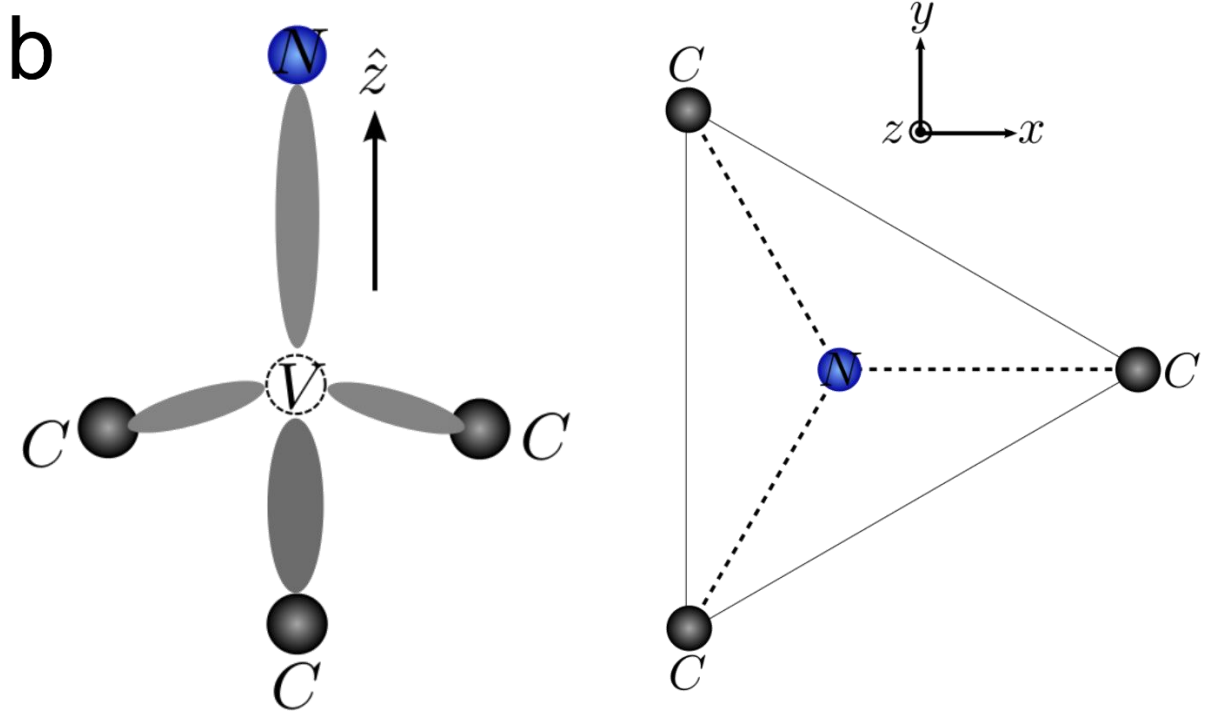

Fig. 7

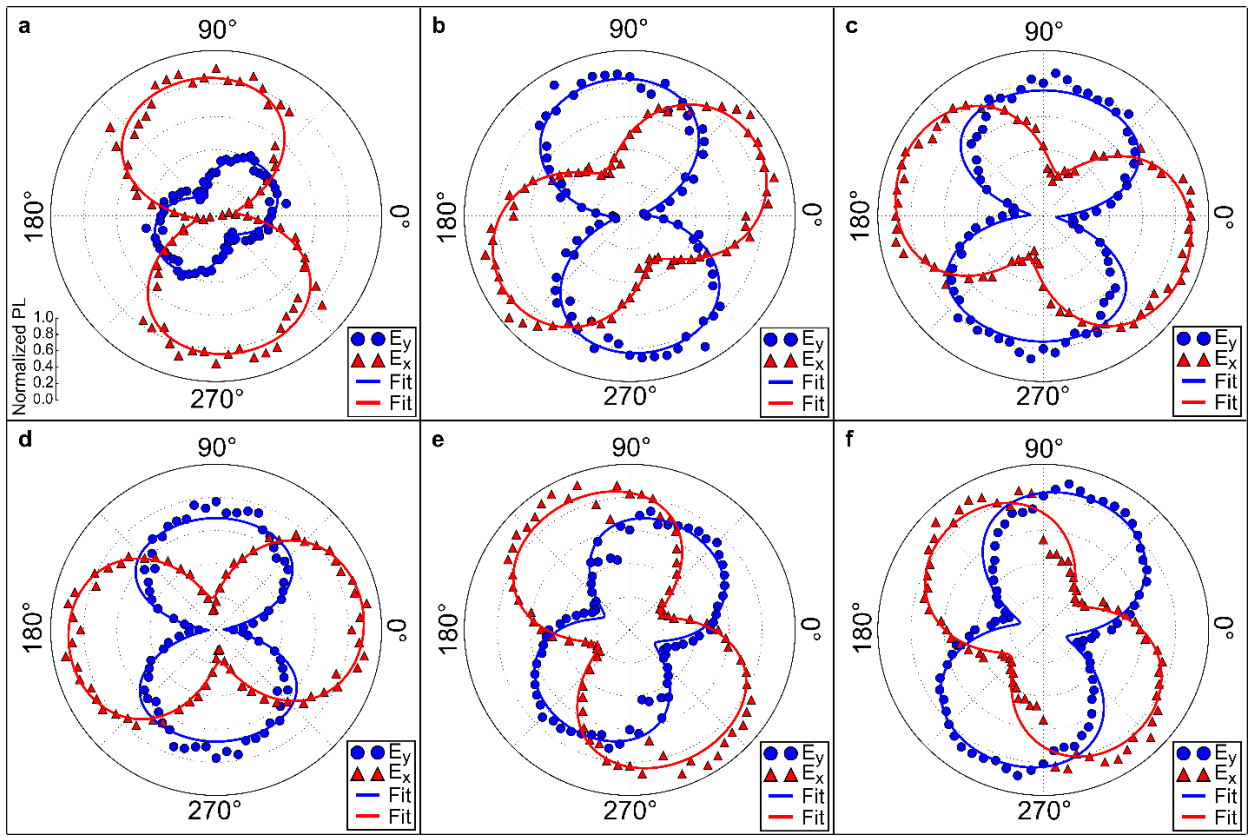

Fig. 8

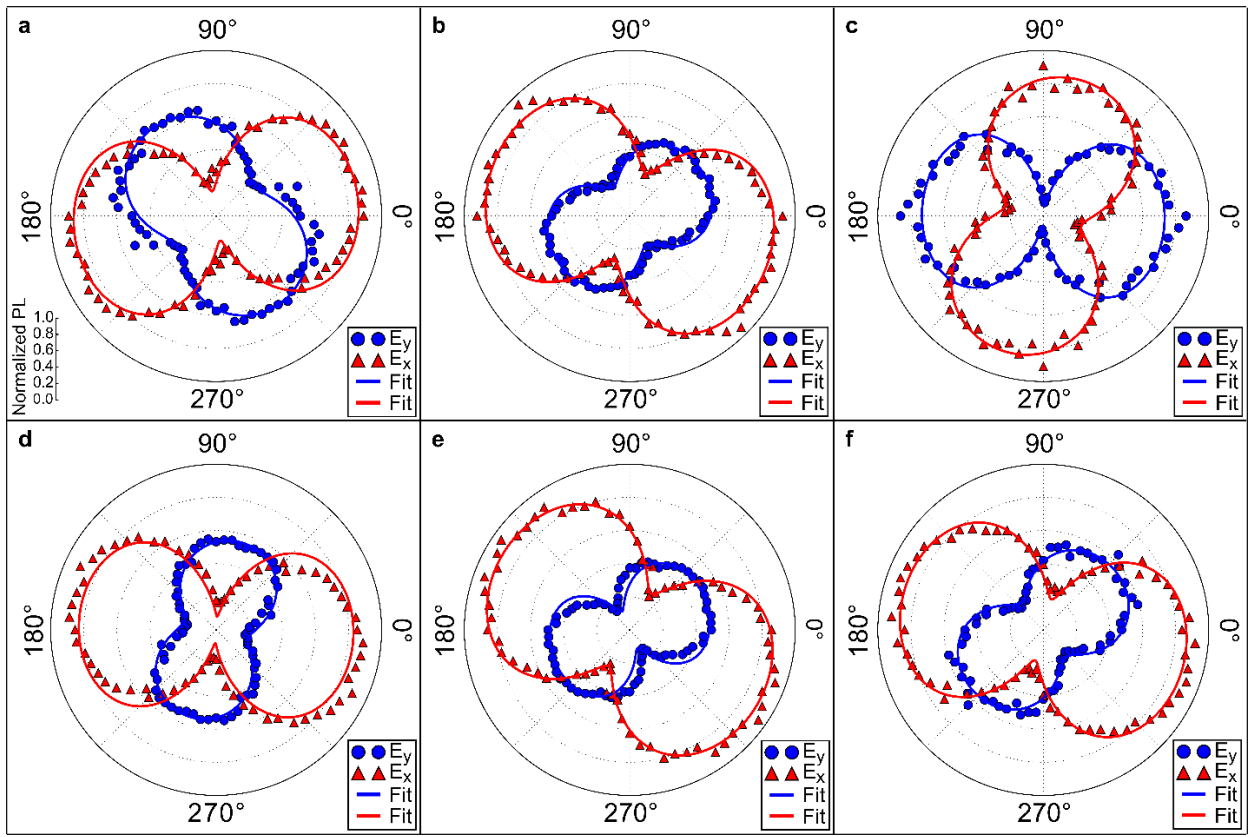

Fig. 9

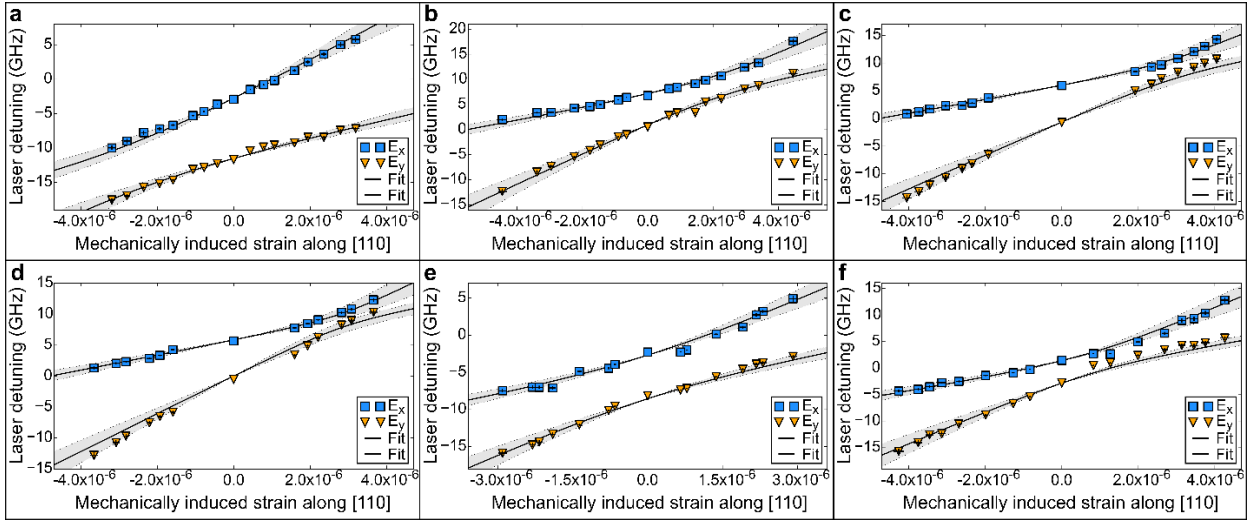

Fig. 10

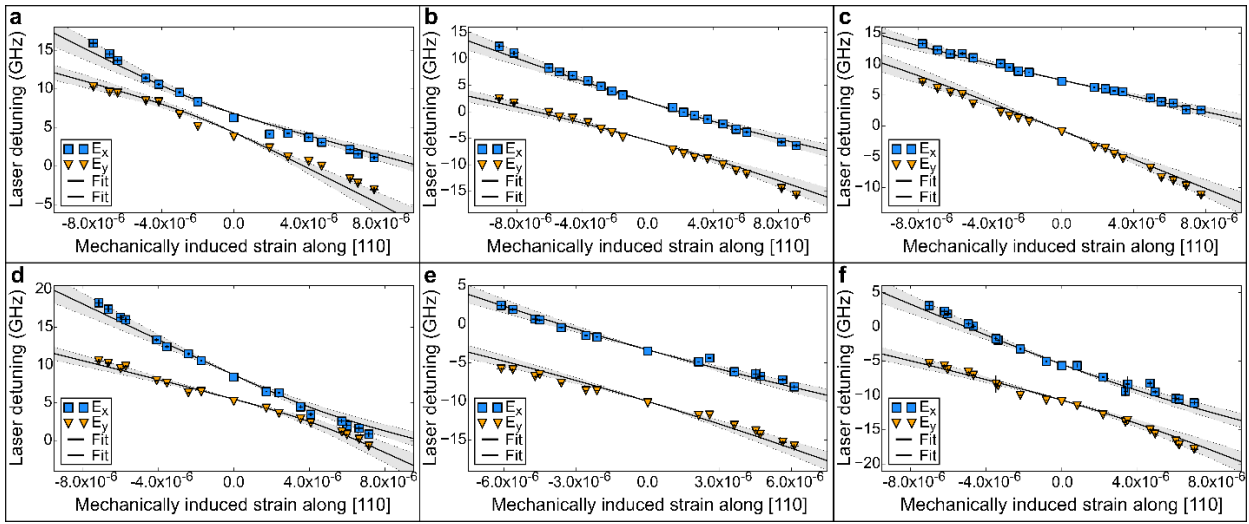

Fig. 11